\documentclass[aps,prd,superscriptaddress,preprint,tightenlines,nofootinbib]{revtex4}
\def\myfigsz{\columnwidth}

\usepackage{graphicx} 
\usepackage{dcolumn}  
\usepackage{bm}       

\usepackage{latexsym}
\usepackage{amsmath}
\usepackage{xspace}
\begin{document}

\preprint{CLNS 07/2011}       
\preprint{CLEO 07-15}         

\title{Measurement of the Absolute Branching Fraction of
$\bm{D^+_s \rightarrow \tau^+ \nu_\tau}$ Decay}

\author{K.~M.~Ecklund}
\affiliation{State University of New York at Buffalo, Buffalo, New York 14260, USA}
\author{W.~Love}
\author{V.~Savinov}
\affiliation{University of Pittsburgh, Pittsburgh, Pennsylvania 15260, USA}
\author{A.~Lopez}
\author{H.~Mendez}
\author{J.~Ramirez}
\affiliation{University of Puerto Rico, Mayaguez, Puerto Rico 00681}
\author{J.~Y.~Ge}
\author{D.~H.~Miller}
\author{I.~P.~J.~Shipsey}
\author{B.~Xin}
\affiliation{Purdue University, West Lafayette, Indiana 47907, USA}
% ...
\collaboration{CLEO Collaboration}
\noaffiliation


%
\date{December 7, 2007}

\begin{abstract}
Using a sample of tagged $D^{+}_s$ decays collected near the
$D^{\ast \pm}_s D^\mp_s$ peak production energy in $e^+ e^-$ collisions
with the CLEO-c detector,
we study the leptonic decay
$D^+_s \rightarrow \tau^+ \nu_\tau $ via the decay channel
$\tau^+ \rightarrow e^+ \nu_e \bar{\nu}_\tau$.
We measure
$\mathcal{B} (D^+_s \rightarrow \tau^+ \nu_\tau)
  =  (6.17 \pm 0.71 \pm 0.34) \%$,
where the first error is statistical and the second systematic.
Combining this result with our measurements
of $D^+_s \rightarrow \mu^+ \nu_\mu$
and
$D^+_s \to \tau^+ \nu_\tau$ (via $\tau^+ \to \pi^+ \bar{\nu}_\tau$),
we determine $f_{D_s}=(274 \pm 10 \pm 5)$ MeV.
\end{abstract}

\pacs{13.20.Fc}

\maketitle

%
In the Standard Model (SM), the decay rate
of a pseudoscalar meson $P_{Q\bar{q}}$ to
a lepton neutrino pair $\ell^+\nu_\ell$ is given by
\begin{equation}
\Gamma (P_{Q\bar{q}} \rightarrow \ell^+ \nu_\ell)
=
\frac{G^2_F |V_{Qq}|^2 f^2_{P}}{8\pi}
     m_{Q\bar{q}} m^2_\ell
     \left(1 - \frac{m^2_\ell}{m^2_{Q\bar{q}}} \right)^2
,
\label{eq:f}
\end{equation}
where
$G_F$ is the Fermi coupling constant,
$V_{Qq}$ is the Cabibbo-Kobayashi-Maskawa (CKM) matrix element,
$m_{Q\bar{q}}$ is the mass of the meson,
and $m_\ell$ is the mass of the charged lepton.
Because no strong interactions are present in the leptonic
final state $\ell^+ \nu_\ell$, such decays provide a clean way to
probe the complex, strong interactions that bind the quark and
anti-quark within the initial-state meson. In these decays, strong
interaction effects can be parametrized by a single quantity, $f_P$,
the pseudoscalar meson decay constant. In the case of the $D^+_s$
meson, $f_{D_s}$ describes the amplitude for the $c$- and $\bar{s}$-quarks
within the $D^+_s$ to have zero separation, a condition necessary for
them to annihilate into the virtual $W^+$ boson that produces the
$\ell^+ \nu_\ell$ pair.

The experimental determination of decay constants is one of the
most important tests of calculations involving non-perturbative
QCD. Such calculations have been performed using various
models~\cite{Richman:1995wm,Penin:2001ux,Narison:2001pu,Ebert:2006hj}
or using lattice QCD~\cite{Aubin:2005ar,Follana:2007uv} (LQCD).
Trustworthy QCD calculations within the $B$-meson sector
would enable the extraction of $|V_{td}|$ from measurements of
$B^0 - \bar{B}^0$ mixing, and $|V_{ub}|$ from
(the very difficult~\cite{Ikado:2006un,Aubert:2005du}) measurements
of $B^+ \to \tau^+ \nu_\tau$. Precision measurements of the decay
constants $f_D$ and $f_{D_s}$ from charm meson decays are an attractive
way to validate the QCD calculations used in the $B$-meson sector.

Physics beyond the SM might also affect leptonic decays of charmed mesons.
Depending on the non-SM features, the ratio of
$\Gamma (D^+ \to \ell^+ \nu_\ell) / \Gamma (D^+_s \to \ell^+ \nu_\ell)$
could be affected~\cite{Akeroyd:2007eh}, as could the ratio
$\Gamma (D^+_s \to \tau^+ \nu_\tau) / \Gamma (D^+_s \to \mu^+ \nu_\mu)$~\cite{Hewett:1995aw,Hou:1992sy}.
Any of the individual widths might be increased or decreased.
In particular, a two-Higgs doublet model~\cite{Akeroyd:2007eh}
predicts a reduction in
$\Gamma (D^+_s \to \ell^+ \nu_\ell)$.

Among the leptonic decays in the charm-quark sector,
$D^+_s \to \ell^+ \nu_\ell$ decays are the most accessible as they are
Cabibbo favored ($|V_{cs}| \sim 1$).
Furthermore, the large mass of the $\tau$ lepton
removes the helicity suppression that is present in the decays to
lighter leptons.
The existence of multiple neutrinos
in the final state, however, makes experimental measurement of this decay
challenging.

In this Letter, we report the most precise
measurement of the absolute branching fraction of the
leptonic decay $D^+_s \to \tau^+ \nu_\tau$,
from which we extract the decay constant $f_{D_s}$
using Eq.~\ref{eq:f}.
%
%
We use a
data sample of
$e^+ e^- \to D^{\ast \pm}_s D^\mp_s$
events
collected by the CLEO-c
detector~\cite{Briere:2001rn,Kubota:1991ww,cleoiiidr,cleorich}
at the center-of-mass (CM)
energy $4170$ MeV, near
$D^{\ast \pm}_s D^\mp_s$  peak production~\cite{Poling:2006da}.
The data sample consists of an
integrated luminosity of $298$ $\text{pb}^{-1}$
provided by the Cornell Electron Storage Ring (CESR).
We have previously reported~\cite{Artuso:2007zg}
measurements of $D^+_s \to \mu^+ \nu_\mu$ and 
$D^+_s \to \tau^+ \nu_\tau$ (via $\tau^+ \to \pi^+ \bar{\nu}_\tau$)
with these data.

%
%
From the interaction point (IP) out,
the CLEO-c
detector~\cite{Briere:2001rn,Kubota:1991ww,cleoiiidr,cleorich}
consists of a six-layer
vertex drift chamber,
a 47-layer central drift chamber,
a ring-imaging Cherenkov detector (RICH),
and a CsI electromagnetic calorimeter,
all operating in a $1.0$ T magnetic field provided by
a superconducting solenoidal magnet.
The detector provides
acceptance of $93 \%$ of the full $4 \pi$ solid angle
for both charged particles and photons.
Charged kaons and pions
are identified based on information from the RICH detector
and the specific ionization ($dE/dx$) measured by the drift chamber.
Electron identification is based on a likelihood variable that combines
the information from RICH detector, $dE/dx$,
and the ratio of electromagnetic shower energy to track momentum ($E/p$).
Background processes and the efficiency of signal-event selection
are estimated with  a GEANT-based~\cite{geant}
Monte Carlo (MC) simulation program.
Physics events are generated by EvtGen~\cite{evtgen},
and
final-state radiation (FSR) is modeled by
the PHOTOS~\cite{photos} program.
The modeling of initial-state radiation (ISR) is based on cross sections
for $D^{\ast \pm}_s D^\mp_s$
production at lower energies obtained
from the CLEO-c energy scan~\cite{Poling:2006da}
near the CM energy where we collect the sample.

%
%
The presence of two $D^\mp_s$ mesons in a $D^{\ast \pm}_s D^\mp_s$ event
allows us to define a single-tag (ST) sample in which a $D^\mp_s$ is
reconstructed in a hadronic decay mode and a further double-tagged (DT)
sub-sample in which an additional $e^\pm$ is required as a
signature of leptonic decay, the $e^\pm$ being the daughter of of
the $\tau^\pm$. The $D^-_s$ reconstructed in the ST sample can either be
primary or secondary from $D^{\ast -}_s \to D^-_s \gamma$
(or $D^{\ast -}_{s} \to \pi^0 D^-_s$).
(We also use charge-conjugate $D_s^+$
decays for the tag; in this Letter, mention of a particular charge
also implies use of the opposite one.)
The ST yield can be expressed as
$n_\text{ST} = 2 N \mathcal{B}_\text{ST} \epsilon_\text{ST}$,
where
$N$ is the produced number of
$D^{\ast \pm}_s D^\mp_s$ pairs,
$\mathcal{B}_\text{ST}$ is the branching fraction of hadronic
modes used in the ST sample,
and $\epsilon_\text{ST}$ is the ST efficiency.

Our double-tag (DT) sample is formed from events with only a single charged
track, identified as a positron, in addition to an ST.
The yield
can be expressed as
$n_\text{DT} = 2 N \mathcal{B}_\text{ST} \mathcal{B}_\text{SG}
\epsilon_\text{DT}$,
where
$\mathcal{B}_\text{SG}$ is the signal decay (SG) branching fraction,
$\epsilon_\text{DT}$ is the efficiency of finding the ST and the SG
in the same event.
From the ST and DT yield expressions we obtain
$\mathcal{B}_\text{SG} =
(n_\text{DT} / n_\text{ST})
\times
(\epsilon_\text{ST} / \epsilon_\text{DT})
=
(n_\text{DT}/\epsilon) / n_\text{ST}
$,
where
$\epsilon$ ($\equiv \epsilon_\text{DT} / \epsilon_\text{ST}$)
is the effective signal efficiency.
Since
$\epsilon_\text{DT} \approx \epsilon_\text{ST} \times \epsilon_\text{SG}$
(where $\epsilon_\text{SG}$ is the SG efficiency),
$\mathcal{B}_\text{SG}$ is nearly independent of the uncertainties
in $\epsilon_\text{ST}$.

%
%
%
To minimize systematic uncertainties,
we tag using three two-body hadronic decay modes with only charged particles
in the final state.
The three ST modes are
$D^-_s \to \phi \pi^-$,
$D^-_s \to K^- K^{\ast 0}$,
and
$D_s^- \to K^0_S K^-$.
The $K^0_S \to \pi^+ \pi^-$ decay is reconstructed by combining
oppositely charged tracks that originate from a common vertex
and that have an invariant mass within $\pm 12$ MeV of the
nominal mass~\cite{Yao:2006px}.
We require the
resonance decay to satisfy the following mass windows around the
nominal mass~\cite{Yao:2006px}:
$\phi \to K^+ K^-$ ($\pm 10$ MeV)
and
$K^{\ast 0} \to K^+ \pi^-$ ($\pm 75$ MeV).
We require the momenta of charged particles to be
$100$ MeV or greater to suppress the slow pion background from
$D^\ast \bar{D}^\ast$ decays (through $D^\ast \to \pi D$).
We identify an ST by using
the invariant mass of the tag $M(D_s)$
and recoil mass against the tag $M_\text{recoil}(D_s)$.
The recoil mass is defined as
$M_\text{recoil}(D_s) \equiv
[ (E_{ee} - E_{D_s} )^2 - |{\bf p}_{ee} - {\bf p}_{D_s}|^2 ]^{1/2}
$,
where $(E_{ee}, {\bf p}_{ee})$ is the net four-momentum of the $e^+ e^-$ beam,
taking the finite beam crossing angle into account;
$(E_{D_s},{\bf p}_{D_s})$ is the four-momentum of the tag,
with $E_{D_s}$ computed from ${\bf p}_{D_s}$ and
the nominal mass~\cite{Yao:2006px} 
of the $D_s$ meson.
We require the recoil mass to be
within $55$ MeV of the $D^\ast_s$ mass~\cite{Yao:2006px}.
This loose window allows both
primary and secondary $D_s$ tags to be selected.

\begin{figure}
  \centering
\includegraphics[width=\myfigsz]{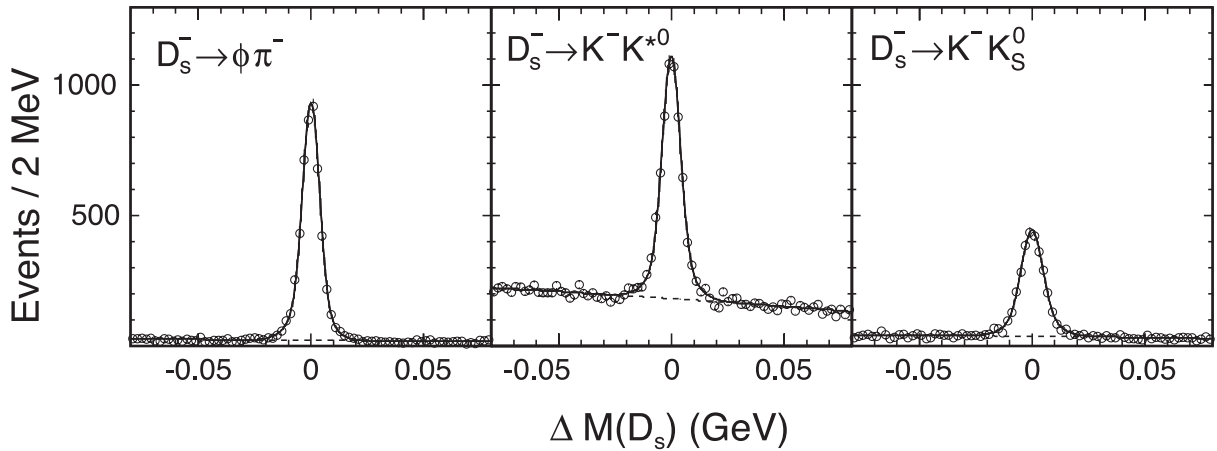}
  \caption{\label{fig:data-fit-dm-8-1-3}
    The mass difference $\Delta M (D_s) \equiv M(D_s) - m_{D_s}$
    distributions in each tag mode.
    We fit the $\Delta M(D_s)$ distribution (open circle)
    to the sum (solid curve)
    of signal (double Gaussian)
    plus background (second degree polynomial, dashed curve)
    functions.
  }
\end{figure}

To estimate the backgrounds
in our ST and DT yields from the wrong tag combinations,
we use the tag invariant mass sidebands.
We define the signal region as 
$-20$ MeV $\le \Delta M(D_s) < +20$ MeV, and the
sideband regions as
$-55$ MeV $\le \Delta M(D_s) < -35$ MeV
or  $+35$ MeV $\le \Delta M(D_s) < +55$ MeV,
where $\Delta M(D_s) \equiv M(D_s) - m_{D_s}$
is the difference between the tag mass and the nominal mass.
We fit the ST $\Delta M(D_s)$ distributions
to the sum of double-Gaussian signal plus second-degree polynomial
background functions to get the sideband scaling factor,
and use that scaling factor for DT events also.
The invariant mass distributions of tag candidates
for each tag mode are shown in Fig.~\ref{fig:data-fit-dm-8-1-3}.

The DT event should have an ST,
a single positron ($p_e \ge 200$ MeV) with no other charged particles, and the
net charge ($Q_\text{net}$) of the event is required to be zero.
These DT events will contain the sought-after
$D^+_s \to \tau^+ \nu_\tau$ ($\tau^+ \to e^+ \nu_e \bar{\nu}_\tau$) events,
but also some backgrounds.
The most effective discrimination variable
that can separate
signal from background events is the extra energy ($E_\text{extra}$)
in the event, \textit{i.e.}, the total energy of the rest of the event.
This quantity is computed
using the neutral shower energy in the calorimeter, counting all
neutral clusters consistent with being photons above $30$ MeV;
these showers must not be associated with any of the ST decay
tracks or the signal positron.
We obtain $E_\text{extra}$ in the signal and sideband regions
of $\Delta M(D_s)$. The sideband-subtracted $E_\text{extra}$
distribution is used to obtain the DT yield.

\begin{figure}
  \centering
  \includegraphics[width=\myfigsz]{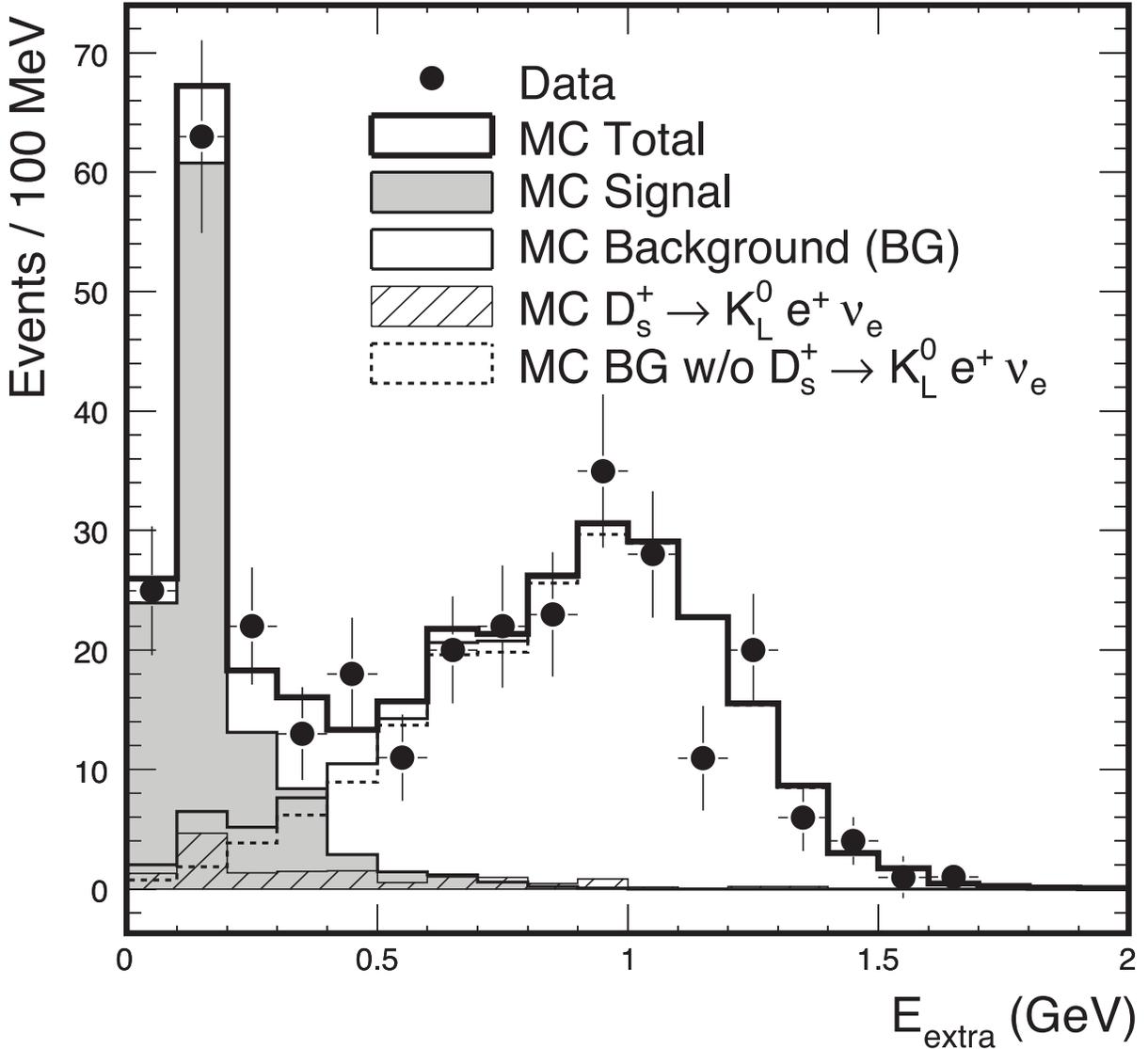}
  \caption{\label{fig:ecc-8-9}
    Distribution of $E_\text{extra}$ after $\Delta M(D_s)$
    sideband subtraction.
  Filled circles are from data and
  histograms are obtained from MC simulation.
  MC signal and the peaking background ($D^+_s \rightarrow K^0_L e^+ \nu_e$)
  components are normalized to our measured branching fractions.
}
\end{figure}

The $E_\text{extra}$ distribution obtained from data is
compared to the MC expectation in Fig.~\ref{fig:ecc-8-9}.
We have used the invariant mass
sidebands, defined above, to subtract the combinatorial background.
We expect that there will be a large peak between
$100$ MeV and $200$ MeV from $D_s^*\to \gamma D_s$ decays
(and from $D_s^* \to \pi^0 D_s$,
$5.8 \%$ branching fraction~\cite{Yao:2006px}).
Also, there will be some events at lower energy when the photon
from $D_s^*$ decay escapes detection.

After the $\Delta M(D_s)$ sideband subtraction, two significant
components of background remain.
One is from $D^+_s \rightarrow K^0_L e^+ \nu_e$. If the $K^0_L$ deposits
little or no energy in the calorimeter, this decay
mode has an $E_\text{extra}$ distribution very similar to the signal,
peaking well below $400$ MeV.
The second source, other semielectronic decays,
rises smoothly with increasing $E_\text{extra}$, up to $1$ GeV.
Estimates of these backgrounds
are also shown in Fig.~\ref{fig:ecc-8-9}.
The optimal signal region in $E_\text{extra}$ for DT yield extraction
is predicted from an MC simulation study.
Choosing $E_\text{extra}$ less than $400$ MeV~\footnote{Note that with our
chosen cut of  $E_\text{extra} < 400$ MeV, we are including
$D^+_S \to \tau^+ \nu_\tau \gamma$ as signal.}
maximizes the signal significance.
The number of non-peaking background events in the $E_\text{extra}$
signal region
is estimated from the number of events in the sideband region above
$600$ MeV scaled by the MC-determined ratio $c_b$
of the number of background events in the signal region, $b^\text{(l)}$,
to the number of events in the sideband region, $b^\text{(h)}$.
The number of peaking background events due to the
$D^+_s \rightarrow K^0_L e^+ \nu_e$ decay
is determined by using the expected number from MC simulation.
The overall expected number of background events
in the $E_\text{extra}$ signal
region ($b$) is computed as follows:
$
b = c_{b} b^\text{(h)}(\text{data})
+
b(K^0_L e^+ \nu_e)_\text{MC}
$,
where
$b^\text{(h)}(\text{data})$ is the number of data events
in the $E_\text{extra}$ sideband region and
$b(K^0_L e^+ \nu_e)_\text{MC}$ is the number of background events
due to $D^+_s \rightarrow K^0_L e^+ \nu_e$ as estimated
from our MC simulation.
The branching fraction for Cabibbo-suppressed decay
$D^+_s \rightarrow K^0_L e^+ \nu_e$ has not yet been measured.
We determine this quantity by measuring
$\mathcal{B} (D^+_s \rightarrow K^0_S e^+ \nu_e)
= (0.14 \pm 0.06 \pm 0.01) \%$
using a sample of $38548$
$D^+_s$ decays (more tag modes are used to increase statistics).

%
%
%
\begingroup
\begin{table*}
\centering
\caption{\label{table:data-double}Summary of
ST yield ($n_\text{ST}$),
ST mass sideband scaling factor ($s$),
DT yield ($n_\text{DT}$) with $400$ MeV cut,
and the number of estimated backgrounds ($b$),
where
$n^S$ is the yield in the ST mass signal region,
and $n^B$ is the yield in the ST mass sideband.}
\begin{ruledtabular}
\begin{tabular*}{\textwidth}{@{\extracolsep{\fill}}l rrrr rr r r}
Tag Mode&
  $n_\text{ST}^S$& $n_\text{ST}^B$& $s$& $n_\text{ST}$&
  $n_\text{DT}^S$& $n_\text{DT}^B$&
    $b$&
    $n_\text{DT}$
\\
\hline
$D_s^- \rightarrow \phi \pi^-$&
$5243$& $391$& $0.997$& $4853.0 \pm 75.1$&
$49$& $0$& $8.8 \pm 0.6$& $40.2 \pm 7.0$
\\
$D_s^- \rightarrow K^- K^{\ast 0}$&
$9020$& $3661$& $1.010$& $5321.0 \pm 112.8$&
$55$& $3$& $8.6 \pm 0.7$& $43.4 \pm 7.6$
\\
$D_s^- \rightarrow K^- K^0_S$&
$3499$& $710$& $1.022$& $2773.1 \pm 65.0$&
$24$& $2$& $4.0 \pm 0.4$& $18.0 \pm 5.1$
\\
\end{tabular*}
\end{ruledtabular}
\end{table*}
\endgroup

The ST yield, $\Delta M(D_s)$ sideband scaling factor,
DT yield with $400$ MeV cut,
and the number of estimated backgrounds
events are summarized in
Table~\ref{table:data-double}.  We find
$n_\text{ST} = 12947 \pm 150$
and
$n_\text{DT} = 102 \pm 12$
integrated over all tag modes.

The signal efficiency determined by MC simulation has been
corrected for a few small differences between data and MC simulation.
We weight the mode-by-mode signal efficiencies 
by the ST yields in each mode to determine
$\epsilon = (71.3 \pm 0.4) \%$ for
the decay chain
$D^+_s \to \tau^+ \nu_\tau \to e^+ \nu_e \bar{\nu}_\tau  \nu_\tau$.
Using
$\mathcal{B} (\tau^+ \rightarrow e^+ \nu_e \bar{\nu}_\tau)
= (17.84 \pm 0.05) \%$~\cite{Yao:2006px}
we obtain the leptonic decay branching fraction
$\mathcal{B} (D^+_s \rightarrow \tau^+ \nu_\tau)
=  (6.17 \pm 0.71) \%$, where the error is statistical.

%
%
The non-positron background in the signal
positron sample is negligible ($0.2\%$)
due to the low probability ($\sim 0.1\%$ per track)
that hadrons ($\pi^+$ or $K^+$) are misidentified as $e^+$.
Uncertainty in these backgrounds produces a
$0.2\%$ uncertainty in the measurement
of $\mathcal{B} (D^+_s \rightarrow \tau^+ \nu_\tau)$.
The secondary positron backgrounds
from charge symmetric processes,
such as $\pi^0$ Dalitz decay ($\pi^0 \to e^+ e^- \gamma$)
and $\gamma$ conversion ($\gamma \to e^+ e^-$), are assessed
by measuring the wrong-sign signal electron in events with
$Q_\text{net} = \pm 2$.
The uncertainty in the measurement from this source is estimated to be
$0.9\%$.
Uncertainties in efficiency due to the extra energy cut ($1.8\%$),
extra track veto ($0.9\%$),
and
$Q_\text{net} = 0$ requirement ($1.3\%$)
are estimated using a sample
in which both  the $D^+_s$ and $D^-_s$ in the event
are tagged with any of the three hadronic ST modes.

We considered five semileptonic decays,
$D^+_s \to$
$\phi e^+ \nu_e$,
$\eta e^+ \nu_e$,
$\eta' e^+ \nu_e$,
$K^0 e^+ \nu_e$,
and $K^{\ast 0} e^+ \nu_e$,
as the major sources of background in the $E_\text{extra}$ signal
region. The first two dominate the non-peaking background,
and the fourth (with $K^0_L$) dominates the peaking background.
Uncertainty in the signal yield due to non-peaking background ($0.5\%$)
is assessed by varying the size of the dominant Cabibbo-favored
semileptonic decays 
by the precision with which they are known~\cite{Yao:2006px}.
Imperfect knowledge of $\mathcal{B}(D^+_s \to K^0 e^+ \nu_e)$
gives rise to a systematic uncertainty in our estimate of the amount
of peaking background in the signal region.
This uncertainty comprises two parts. We estimate the $K^0_L$
showering systematic uncertainty using $\psi (3770)$ events in which the
$\bar{D}^0$ has been fully reconstructed in a hadronic mode and the $D^0$
decays as $D^0 \to K^0_L \pi^+ \pi^-$.
When this uncertainty is combined in quadrature with the
uncertainty in the determination of $\mathcal{B} (D^+_s \to K^0_S e^+ \nu_e)$,
the systematic uncertainty on $\mathcal{B} (D^+_s \to \tau^+ \nu_\tau)$ is
$4.5 \%$.

Other possible sources of systematic uncertainty
include
$n_\text{ST}$ ($0.8\%$),
tracking efficiency ($0.3\%$),
positron identification efficiency ($1\%$),
and FSR ($1\%$).
Combining all contributions in quadrature,
the total systematic uncertainty in the branching fraction measurement
is estimated to be
$5.5 \%$.

%
%
In conclusion, using a sample of $D^{+}_s$ decays collected
with the CLEO-c detector,
we obtain a measurement of the absolute branching fraction,
$\mathcal{B} (D^+_s \rightarrow \tau^+ \nu_\tau) =
(6.17 \pm 0.71 \pm 0.34) \%$,
where the first error is statistical and the second is systematic.
This is the most precise measurement of this branching fraction
and does not depend on measurements of other $D_s$ branching fractions
for normalization.
The decay constant
$f_{D_s}$
can be computed using Eq.~\ref{eq:f}
with known values~\cite{Yao:2006px} of
$G_F = 1.16637(1)\times 10^{-5}$ GeV$^{-2}$,
$|V_{cs}| = 0.9738$~\footnote{We assume $|V_{cs}| = |V_{ud}|$
and use the value given in Ref.~\cite{Yao:2006px}.},
$m_{D_s} = 1968.2(5)$ MeV,
$m_\tau = 1776.99^{+0.29}_{-0.26}$ MeV,
and the lifetime of $\tau_{D_s} = 500(7) \times 10^{-15}$ s
(errors from these input parameters are negligible and ignored).
We obtain
$f_{D_s} =  (273 \pm 16 \pm 8)$ MeV.
Combining with our previous decay constant
determination~\cite{Artuso:2007zg}
of $f_{D_s} = (274 \pm 13 \pm 7)$ MeV, we obtain
$f_{D_s} = (274 \pm 10 \pm 5)$ MeV.
Our measured decay constant is consistent with the
world average $f_{D_s} = (294 \pm 27)$ MeV~\cite{Yao:2006px}
and another recent measurement 
$f_{D_s} = (283 \pm 17 \pm 7 \pm 14)$ MeV~\cite{Aubert:2006sd}.
These results are generally higher than recent
LQCD calculations
$f_{D_s} = (249 \pm 3 \pm 16)$ MeV~\cite{Aubin:2005ar}
and
$f_{D_s} = (241 \pm 3)$ MeV~\cite{Follana:2007uv}.
The predicted suppression~\cite{Akeroyd:2007eh} that would be caused
by a charged Higgs seems to be incompatible with experimental measurements
combined with LQCD calculations.

Combining with our previous measurement~\cite{Artuso:2007zg}
of
$D^+_s \rightarrow \tau^+ \nu_\tau $
($\tau^+ \to \pi^+ \bar{\nu}_\tau$),
we obtain
$\mathcal{B} (D^+_s \rightarrow \tau^+ \nu_\tau)
  =  (6.47 \pm 0.61 \pm 0.26) \%$.
Using this with our measurement~\cite{Artuso:2007zg}
of $D^+_s \rightarrow \mu^+ \nu_\mu$,
we obtain the branching fraction ratio
$\frac{\mathcal{B}(D^+_s \rightarrow \tau^+ \nu_\tau)}
      {\mathcal{B}(D^+_s \to \mu^+ \nu_\mu)} = 11.0 \pm 1.4 \pm 0.6$.
This is consistent with $9.72$, the value predicted by the SM
with lepton universality~\cite{Hewett:1995aw,Hou:1992sy},
as given in Eq.~\ref{eq:f}.

We gratefully acknowledge the effort of the CESR staff
in providing us with excellent luminosity and running conditions.
This work was supported by
the A.P.~Sloan Foundation,
the National Science Foundation,
the U.S. Department of Energy,
the Natural Sciences and Engineering Research Council of Canada, and
the U.K. Science and Technology Facilities Council.

\end{document}